%% file: ReCAFR.tex
\documentclass[sigconf]{acmart}
\AtBeginDocument{%
  }
\usepackage{multirow}
\copyrightyear{2025}
\acmYear{2025}
\setcopyright{cc}
\setcctype[4.0]{by}
\acmConference[WSDM '25]{Proceedings of the Eighteenth ACM International Conference on Web Search and Data Mining}{March 10--14, 2025}{Hannover, Germany}
\acmBooktitle{Proceedings of the Eighteenth ACM International Conference on Web Search and Data Mining (WSDM '25), March 10--14, 2025, Hannover, Germany}\acmDOI{10.1145/3701551.3703530}
   
\acmISBN{979-8-4007-1329-3/25/03}

\usepackage{enumitem}
\usepackage{algorithm}
\usepackage{algpseudocode}
\newcommand{\stitle}[1]{\vspace{1mm} \noindent {\bf #1}}
\renewcommand{\vec}[1]{\ensuremath{\mathbf{#1}}}

\usepackage{colortbl}
\definecolor{verylightgray}{rgb}{0.9, 0.9, 0.9} 

\begin{document}

\title{A Contrastive Framework with  User, Item and Review Alignment for Recommendation}


\author{Hoang V. Dong}
\affiliation{%
    \institution{Singapore Management University}
    \country{Singapore}
}
\email{vhdong.2021@smu.edu.sg}
\author{Yuan Fang}
\affiliation{%
    \institution{Singapore Management University}
    \country{Singapore}
}
\email{yfang@smu.edu.sg}  
\author{Hady W. Lauw}
\affiliation{%
    \institution{Singapore Management University}
    \country{Singapore}
}
\email{hadywlauw@smu.edu.sg}


\input{container/_abs}




\maketitle

\input{container/_intro}
\input{container/_relatedWorks}
\input{container/_Preliminaries}
\input{container/_approach}
\input{container/_results}
\input{container/_conclusion}

\section*{Acknowledgments}
This research / project is supported by the Ministry of Education, Singapore, under its Academic Research Fund Tier 2 (MOE-T2EP20122-0041). Any opinions, findings and conclusions or recommendations expressed in this material are those of the author(s) and do not reflect the views of the Ministry of Education, Singapore.

\newpage

\section*{Ethical Statement}
Our work aims to improve the effectiveness of recommendation systems.
From a social perspective, more effective personalized recommendations can significantly enhance the visibility and accessibility of underrepresented groups. These groups may benefit from increased exposure of their content or voices to a broader audience; users from varied backgrounds may enjoy more accessible and inclusive offerings.
While our work does not present immediate or specific ethical concerns, the broader application of machine learning and artificial intelligence techniques carries inherent risks.

\balance
\bibliographystyle{ACM-Reference-Format}
\bibliography{container/references}
\newpage
\appendix
\input{container/_appendix}

\end{document}

%% file: container/_abs.tex
\begin{abstract}
Learning effective latent representations for users and items is the cornerstone of recommender systems. Traditional approaches rely on user-item interaction data to map users and items into a shared latent space, but the sparsity of interactions often poses challenges. While leveraging user reviews could mitigate this sparsity, existing review-aware recommendation models often exhibit two key limitations. First, they typically rely on reviews as additional features, but reviews are not universal, with many users and items lacking them. Second, such approaches do not integrate reviews into the user-item space, leading to potential divergence or inconsistency among user, item, and review representations.
To overcome these limitations, our work introduces a Review-centric Contrastive Alignment Framework for Recommendation (ReCAFR), which incorporates reviews into the core learning process, ensuring alignment among user, item, and review representations within a unified space. Specifically, we leverage two self-supervised contrastive strategies that not only exploit review-based augmentation to alleviate sparsity, but also align the tripartite representations to enhance robustness. Empirical studies on public benchmark datasets demonstrate the effectiveness and robustness of ReCAFR.
\end{abstract}

%% file: container/_intro.tex
\section{Introduction}\label{sec:intro}
Personalized recommendations are prevalent in diverse online services such as social networks \cite{konstas2009social}, e-commerce \cite{schafer2001commerce}, and advertising \cite{zhou2020product}. Their key idea is to predict user-item interactions like clicks and ratings based on observed interaction data. This technique, known as collaborative filtering \cite{breese2013empirical,ekstrand2011collaborative}, operates on the assumption that users exhibiting similar behavioral patterns are likely to share comparable preferences for various items. 
Both classical and contemporary methods, such as matrix factorization \cite{bpr} and LightGCN \cite{4}, learn a shared latent space for all users and items based on observed interaction data, so that users and items conform to the collaborative assumption in the shared space.


\stitle{Prior work.} 
Collaborative filtering methods often encounter the challenge of data sparsity \cite{bpr} in user-item interactions. An intuitive and popular solution to address sparsity involves leveraging additional side information, such as user and item attributes \cite{li2017collaborative,zhu2019addressing}, reviews \cite{NARRE} and social factors \cite{fan2019graph,tang2013social}. In particular, review data 
have emerged as a valuable source of information, owning to their rich semantic content. Written by diverse users for a wide range of items, reviews often encompass contextual information for grasping both user preferences and item characteristics. Such contextual details could be utilized for more accurate and personalized recommendations, as they go beyond binary interactions or numerical ratings to include qualitative and multi-aspect insights. 

 Given the semantic richness of reviews, numerous studies \cite{zheng2017joint,tay2018multi,NARRE,wu2018parl, RGCL, ANCF,SGDN,RMCL,ReHCL} have integrated review data into collaborative filtering. Existing review-aware methods typically treat reviews as additional user or item features to enrich their representations. 
For instance, NARRE \cite{5} aggregates the pool of reviews of each user or item using an attention mechanism, 
which is then combined with the latent factors of the users and items to obtain their final representations. In another study, RGCL \cite{RGCL} employs the reviews as edge features on a bipartite user-item graph, where each edge denotes an interaction between user and item. 

\stitle{Challenges.} However, directly utilizing reviews as features may not fully exploit the potential of review data. First, although prevalent on online platforms, reviews are far from universally available, as many users or items have no reviews. 
Thus, directly deriving features from the reviews would lead to many users or items with missing features, necessitating padding or imputation techniques, which could adversely impact performance. 
Second, relying on reviews merely as features treats them as supplementary to the user-item latent space, 
which can lead to inconsistencies between the representations of a review and its associated user or item, thereby undermining the effectiveness of collaborative filtering. 
Hence, to fully exploit the potential of review data, we must address two open research questions. 
(1) \emph{How can we seamlessly integrate reviews into the collaborative filtering models}, rather than merely employing them at the feature level? 
Review data comprise unstructured text, which are inherently different from interaction data, posing a challenge in integrating them. Moreover, many items and users do not associate with any review, simply using reviews as features may result in significant missing features in real-world applications. 
(2) \emph{How can we align user, item and review representations in a shared latent space} to ensure their consistency? Reviews encompass semantic signals that may exhibit considerable discrepancies with the collaborative signals derived from interaction data. Thus, aligning the tripartite representations is essential to achieve synergy between the two types of signals.


\stitle{Present work.} To address these questions, we propose the \textbf{Re}view-centric \textbf{C}ontrastive \textbf{A}lignment \textbf{F}ramework for \textbf{R}ecommendation (ReCAFR).
It integrates both reviews and interaction data 
into a contrastive framework, and further aligns the tripartite representations for users, items, and reviews within a unified space. 
Notably, ReCAFR operates on a self-supervised formulation, requiring no extra annotation. 
Self-supervised learning 
has gained popularity in related domains including computer vision \cite{unsupervised,cv2} and natural language processing \cite{bert,albert}. While there are quite a number of attempts at self-supervised recommendation \cite{RGCL,sgl,wang2023sequential,xie2022contrastive}, they focus on graph- \cite{RGCL,sgl} or sequence-based \cite{wang2023sequential,xie2022contrastive} augmentation, but we focus on review-centric augmentation.

\begin{figure}[t]
\centering
\includegraphics[width=0.99\columnwidth]{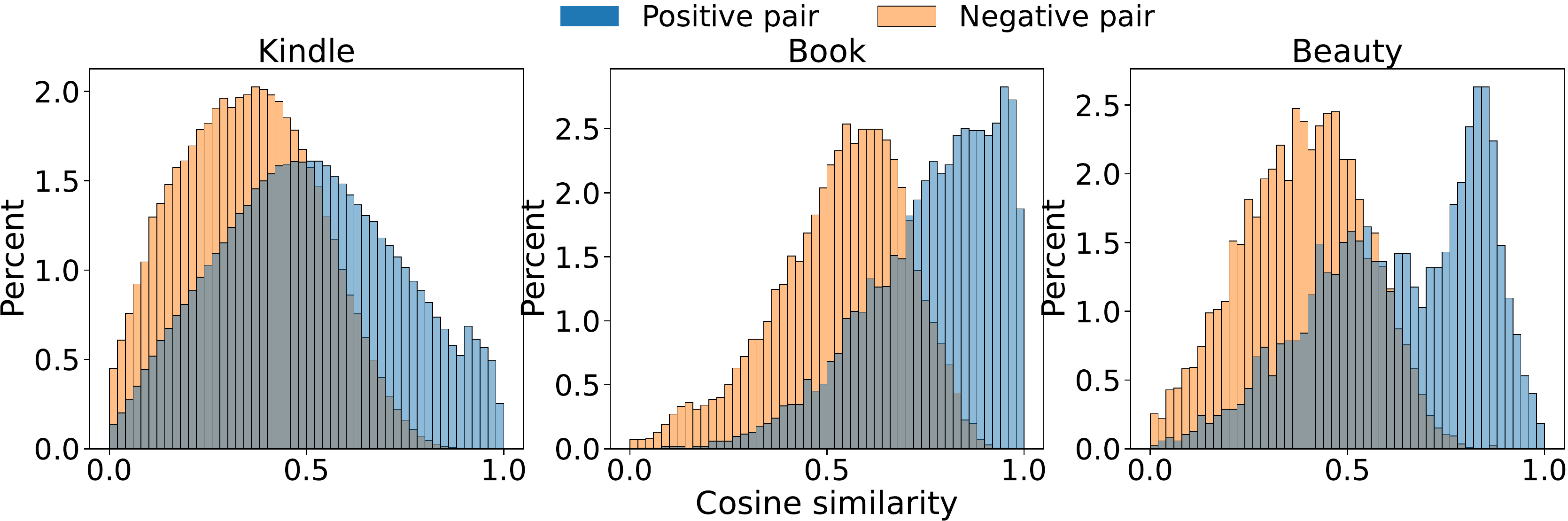}
\vspace{-2mm}
\caption{Comparison of the similarity distribution of the positive pairs versus that of the negative pairs.}
\label{fig:pilot}
\end{figure}

More specifically, toward the first research question, we employ review data for augmentation within a contrastive learning framework to alleviate the sparsity of interaction data. We observe that the same user typically exhibits consistent reviewing patterns, such as writing style and focus on certain aspects, in contrast to the divergent patterns displayed by different users. These patterns, effectively serving as a unique ``user signature,'' emerge because reviews implicitly capture the underlying user preferences.
To leverage this observation, we design a self-supervised task that contrasts reviews from the same or different users. 
Let $S^k_x$ represent the $k$th sampled view of reviews written by the user $x$. Our proposed contrastive strategy aims to discriminate between a pair $(S^1_u, S^2_{u})$, i.e., two views of reviews originating from the same user $u$, and a pair $(S^1_u, S^2_{u'})$, i.e., two views of reviews originating from two distinct users $u\ne u'$.
In the contrastive framework, $(S^1_u, S^2_{u})$ is classified as a positive pair, whereas $(S^1_u, S^2_{u'})$ is regarded as a negative pair. In Fig.~\ref{fig:pilot}, we present an analysis comparing the similarity distribution of the positive pairs, and that of the negative pairs constructed based on the reviews across three Amazon datasets\footnote{See Sect.~\ref{sect:expt:setup} for details of the datasets, and 
Appendix A for details of the calculation.}. The comparison shows that the positive pairs are more likely to be similar, indicating consistent review patterns of the same user. Conversely, the negative pairs are less likely to be similar, reflecting distinct review patterns or preferences among different users. That is, reviews from the same user tends to be more similar than reviews from a different user. 
A parallel observation can be made from the item perspective: reviews pertaining to the same item tend to exhibit greater similarity compared to those of different items, suggesting that a similar contrastive setup can be employed. 

To address the second research question, we propose another contrastive strategy to align the user, item and review-based representations. 
For each user $u$, on one hand, we derive a vector representation $\vec{e}_u$ from the collaborative signals in the interaction data. On the other hand, we sample a view of the reviews written by each user $u$, and derive a corresponding vector $\vec{h}_u$ from the semantic signals in the review texts of the sampled view. While both  $\vec{e}_u$ and $\vec{h}_u$ aim to encode the underlying preferences of $u$, they lie in separate latent spaces and may display inconsistencies. 
We seek to maximize the agreement between $\vec{e}_u$ and $\vec{h}_u$ to achieve consistent user-review representations from both collaborative and semantic perspectives. 
A similar alignment can be performed on the item side as well, to maximize the agreement between the collaborative embedding $\vec{e}_i$ and semantic embedding $\vec{h}_i$. These alignment tasks ensure that the tripartite  representations for users, items, and reviews converge to a shared latent space, thereby enhancing the robustness of the model.

\stitle{Contributions.}
The main contributions of this work are threefold.
(1) We identify the limitations inherent in treating review data merely as features, and observe that reviews provide distinctive contrastive signals for both user and item sides.
(2) We propose ReCAFR, a Review-centric Contrastive Aligment Framework for Recommendation. ReCAFR not only employs review data for augmentation to mitigate the sparsity problem, but also aligns the tripartite representations to improve robustness.
(3) We conduct extensive experiments on real-word datasets to demonstrate the effectiveness of ReCAFR.

%% file: container/_relatedWorks.tex
\section{Related Work}
In this section, we review literature for recommendation systems, including review-aware and self-supervised methods.

\stitle{Collaborative filtering.}
Matrix factorization (MF) is a classical and widely used approach in collaborative filtering \citep{bennett2007netflix,koren2009matrix,ricci2015recommender}. It projects users and items into a latent space, aiming to capture latent factors that represent user preferences and item characteristics. 
The latent factors are optimized to fit the user-item interaction matrix.
Recent developments in neural recommender models, such as NCF \citep{NCF} and LRML \citep{tay2018latent}, adhere to the same concept of latent factors, while enhancing the interaction modeling with neural networks. More advanced neural network architectures have also been proposed, including the attention mechanism \citep{ACF,nais} to discern  and weigh the importance of each interacted item, and graph convolutional networks \cite{4} to exploit high-order interactions. Additionally, DirectAU \cite{DirectAU} investigates the desired properties of representations in collaborative filtering. 
Other aspects of collaborative filtering have also been explored, such as cold-start recommendation \cite{lu2020meta}, online learning~\cite{he2016fast} and continual learning~\cite{do2023continual}, and integration with large language models \cite{RLMRec,liu2024collaborative}.

\stitle{Review-aware recommendation.}
Reviews have been a popular source of side information to mitigate the sparsity issue in interaction data \cite{HFT,ling2014ratings,wu2019reviews,wu2019context,liu2019daml,dong2020asymmetrical,gao2020set,RGCL}. 
As reviews are expressed in natural language, text-based techniques are often incorporated into the recommendation model \cite{dong2020asymmetrical,seo2017interpretable,wu2019context,zheng2017joint,dao2024broadening}.
For instance, to extract semantic features for users and items from reviews, 
DeepCoNN \cite{zheng2017joint} employs two concurrent TextCNNs \cite{kim2014convolutional}. NARRE \cite{NARRE} leverages the attention mechanism \cite{vaswani2017attention} to select relevant reviews, which can also enhance the interpretability of the model. Beside, RMCL \cite{RMCL} effectively reduces noise information from the reviews by adopting a multi-intention contrastive strategy.
These methods rely on reviews to derive or extract part of their input features for users, items, or interactions, which can result in missing features when reviews are unavailable. In contrast, our work integrates reviews and collaborative filtering in a unified space, enhancing their synergy and improving robustness.


\stitle{Self-supervised learning for recommendation.}
Autoencoder-based recommendation  models \cite{wu2016collaborative} are earlier examples of self-supervised techniques for recommendation. Self-supervised learning \cite{bert,mikolov2013distributed,hjelm2018learning} aims to 
train a model through an auxiliary task, for which grounth-truth samples can be automatically extracted from the original input without requiring additional annotations. Both generative tasks \citep{bert} and contrastive tasks \citep{contrastive1, unsupervised} are popular choices for self-supervised learning. 

In recommendation systems, BERT4Rec \cite{sun2019bert4rec} applies a language model \cite{bert} for sequential recommendation tasks. By adopting a bidirectional representation model, BERT4Rec improves over the left-to-right training strategy utilized by SASRec \cite{kang2018self}. In these approaches, some items in the input sequences are randomly masked, which are then predicted based on the surrounding items.
Additionally, contrastive methods have gained increasing traction in recommendation systems \cite{yao2021self,tao2022self,guo2022miss}. Their basic principle is to apply data augmentation to an instance (e.g., user, item, or sequence)---variants of the same instance should be more similar than variants of different instances.  For example, SL4Rec \cite{yao2021self} applies feature augmentation including masking and dropout on items, and further employs a two-tower design on a pair of augmented instances for contrastive learning. 
SGL \cite{sgl}, SimGCL\cite{SimGCL} and RGCL \cite{RGCL} exploit graph-based contrastive learning on the user-item graph, leveraging structural augmentations such as node and edge dropout.
Distinct from these works, our contrastive learning applies augmentation on the reviews to extract semantic signals, serving to complement the collaborative signals.  

%% file: container/_Preliminaries.tex
\section{Preliminaries}\label{sec:prelim}

\begin{figure*}[h]
\centering
\includegraphics[width=0.87\textwidth]{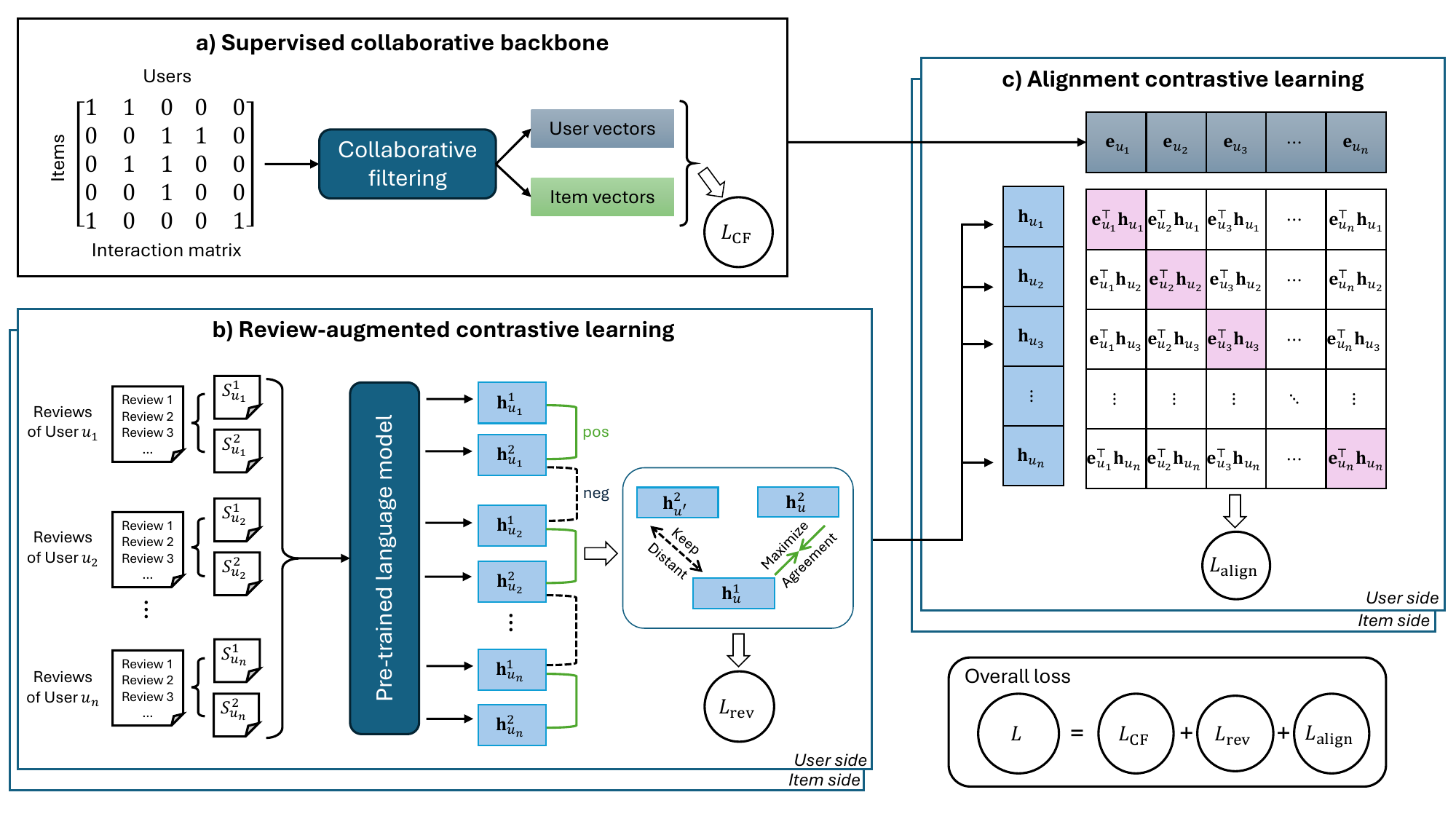} 
\vspace{-2mm}
\caption{Overall framework of ReCAFR. For brevity, it illustrates (b) review-augmented contrastive learning and (c) alignment contrastive learning for the user side only; similar procedures are applied to the item side.}
\label{fig:framework}
\end{figure*}
We first introduce the recommendation problem, and then summarize the conventional collaborative filtering techniques.

\stitle{Problem statement.} 
Consider a set of users $U$ and a set of items $I$. The user-item interaction matrix $Y \in \mathbb{R}^{|U|\times|I|}$ is given by
\begin{align}
    y_{u,i} = 
    \begin{cases}
    1,  & u \text{ has an observed interaction with } i,  \\
    0,  &\text{otherwise.}
    \end{cases}
\end{align}
for any user $u\in U$ and item $i \in I$. What constitutes an interaction between $u$ and $i$ is often application-specific, such as $u$ clicking on, bookmarking, buying, or rating $i$.

Conventional approaches rely solely on the interaction matrix $Y$. To address the sparsity in $Y$, reviews can be leveraged to enrich the interaction data. In our setting, we assume a set of reviews $S_u$ written by user $u$, as well as a set of reviews $S_i$ written for item $i$. A review is regarded as a text document, i.e., $S_u$ or $S_i$ represents a set of documents for $u$ or $i$. In practical scenarios, some users or items may lack any review, implying that $S_u=\emptyset$ or $S_i=\emptyset$ for them. 

\stitle{Collaborative filtering.} 
Our review-augmented contrastive framework can incorporate any collaborative filtering backbone \cite{bpr, 4,sgl,SimGCL, DirectAU}. Here, we briefly summarize matrix factorization \cite{bpr}, a classical collaborative filtering approach.

Matrix factorization associates a user or an item with an embedding vector, which is termed as the latent factors of the user or item. The latent factors aim to capture the user preferences or item characteristics. Let $\mathbf{e}_u \in \mathbb{R}^d$ denote the latent factors of user $u$, and $\mathbf{e}_i \in \mathbb{R}^d$ denote the latent factors of item $i$. Matrix factorization estimates an interaction $y_{u,i}$ as the inner product of the latent factors of $u$ and $i$, i.e.,
$\hat{y}_{u,i} = \mathbf{e}_u^\top \mathbf{e}_i$. 
To optimize the latent factors and other model parameters, existing works usually frame the task as supervised learning based on the observed interactions. In this work, we optimize the pairwise Bayesian Personalized Ranking (BPR) loss \citep{bpr}, which takes into account the relative ordering between positive (observed) and negative (unobserved) user-item interactions. 
The training set $\mathcal{O}$ consists of a series of triples, $\mathcal{O}=\{(u,i,j)\mid y_{u,i}=1, y_{u,j}=0\}$, indicating that in each triple  $(u,i,j)$, user $u$ has interacted with item $i$ but not with item $j$.  The BPR approach postulates that $u$ would favor item $i$ over item $j$, as follows.
\begin{align}\label{eq:loss_cf}
\textstyle L_\text{CF} = \sum_{(u,i,j)\in \mathcal{O}} -\ln \sigma (\hat{y}_{u,i} - \hat{y}_{u,j}), 
\end{align}
where $\hat{y}_{u,i}$ and $\hat{y}_{u,j}$ are estimated using matrix factorization or any collaborative filtering technique.

%% file: container/_approach.tex
\section{Proposed Approach}\label{sec:approach}

In this section, we introduce our approach ReCAFR. We start with an overview of ReCAFR, followed by its concrete realization. 

\subsection{Overview}
Figure~\ref{fig:framework} illustrates the overall framework, which augments the supervised collaborative backbone with self-supervised contrastive learning. 
As depicted in Figure~\ref{fig:framework}(a), the collaborative backbone takes user-item interactions as input, and generates vector representations of users and items using any collaborative filtering technique. In particular, the interaction data act as supervisory signals for optimizing the collaborative filtering loss.

We further leverage review data for review-augmented contrastive learning, as illustrated in Fig.~\ref{fig:framework}(b). A user (or an item) can be associated with a set of reviews, which describe the underlying preferences of the user (or characteristics of the item). For the purpose of contrastive learning, we sample different views from these reviews to formulate contrastive examples in a self-supervised manner. 
Ultimately, to maximize the potential of review data and ensure its alignment with the collaborative signals, we seek to align user and item representations from both the collaborative backbone and the review-augmented contrastive learning within a shared latent space. This alignment is achieved through another contrastive learning module, as illustrated in Fig.~\ref{fig:framework}(c).

In the following, we elaborate the two contrastive learning modules (Sects.~\ref{sec:approach:review-augmented} and \ref{sec:approach:alignment}), as well as the overall training and inference procedures (Sect.~\ref{sec:approach:overallloss}).



\subsection{Review-Augmented Contrastive Learning}\label{sec:approach:review-augmented}

To alleviate the sparsity of interaction data, we incorporate review data into collaborative filtering. As motivated in Sect.~\ref{sec:intro}, for improved robustness against missing reviews, we integrate review data via a contrastive framework, departing from existing approaches that treat reviews as input features \cite{NARRE,RGCL}. 

\stitle{Review augmentation.} We employ reviews for data augmentation on both the user and item sides. The two sides are largely symmetric and we primarily use the user side to exemplify. The underlying principle is that each review, written by a user, implicitly capture the user's preferences. As a result, reviews written by the same user tend to show more similar patterns than reviews by different users, a phenomenon we have demonstrated through the diverging distributions in the preliminary analysis (Fig.~\ref{fig:pilot}). Intuitively, each review embeds a unique ``signature'' of its author.

Specifically, for a user $u$ with a set of reviews $S_u$, we sample two different views from these reviews, designated as $S_u^1$ and $S_u^2$, where $S_u^k\subset S_u$ represents the $k$th view of the reviews written by user $u$. Furthermore, we require that $S_u^1$ and $S_u^2$ be disjoint. The two views $S_u^1$ and $S_u^2$ encapsulate the same inherent signature of $u$, whereas $S_u^1$ 
and $S_{u'}^2$ contain distinct signatures for different users $u$ and $u'$, respectively. This contrast naturally suggests that $(S_u^1,S_u^2)$ forms a positive pair likely with a higher similarity, while $(S_u^1,S_{u'}^2)$ constitutes a negative pair likely with a lower similarity. Subsequently, the positive and negative pairs serve as data augmentation for contrastive learning.

Analogously, reviews for the same item tend to be more similar than those for different items.
For an item $i$ with a set of reviews $S_i$, we sample two views  $S_i^1,S_i^2\subset S_i$ to form a positive pair $(S_i^1,S_i^2)$, while a negative pair $(S_i^1,S_{i'}^2)$ can be constructed using another item $i'\ne i$.

\stitle{Review-augmented contrastive strategy.}
Given the contrasting positive and negative pairs from user and item reviews, we introduce our review-augmented contrastive strategy. 

First, our method employs a pre-trained language model, $f(\cdot)$, to initialize the representation of each review.
Specifically, $f(s)$ maps a review $s$ into a $d$-dimensional vector. Subsequently, we compute a view-level representation for each sampled view. 
In particular,  for the $k$th view of user $u$, $S^k_u$, we compute its representation, $\mathbf{h}^k_{u}$, using mean pooling of the representations of all reviews in the view, with a similar setup at the item side:
\begin{align}
\mathbf{h}^k_{u}=\textsc{Mean}(\{f(s)\mid s\in S^k_u\}),\label{eq:view_embedding_u}\\
\mathbf{h}^k_{i}=\textsc{Mean}(\{f(s)\mid s\in S^k_i\}).\label{eq:view_embedding_i}
\end{align}

Note that the pre-trained langauge model $f(\cdot)$ is only used to \emph{initialize} the view-level representations. 
The initial representations will be further optimized through the contrastive objective.
We adopt the InfoNCE loss \citep{InfoNCE} to maximize the agreement of the positive pairs and minimize that of the negative pairs, on both user and item sides, as follows.
\begin{align}
L^\text{user}_\text{rev} & = \sum_{u\in U}  - \log \frac{\exp(\operatorname{sim}(\mathbf{h}^1_{u},\mathbf{h}^2_{u})/\tau)}{ \sum_{u'\in U}\exp(\operatorname{sim}(\mathbf{h}^1_{u},\mathbf{h}^2_{u'})/\tau)) }, \label{eq:rev_user} \\
L^\text{item}_\text{rev} & = \sum_{i\in I}  - \log \frac{\exp(\operatorname{sim}(\mathbf{h}^1_{i},\mathbf{h}^2_{i})/\tau)}{ \sum_{i'\in I}\exp(\operatorname{sim}(\mathbf{h}^1_{i},\mathbf{h}^2_{i'})/\tau)) }, \label{eq:rev_item}
\end{align}
where $\operatorname{sim}(\cdot)$ measures the similarity between two vectors, implemented as cosine similarity in our experiments, and $\tau$ is the temperature hyperparameter. 

\subsection{Alignment Contrastive Strategy} \label{sec:approach:alignment}

To improve the synergy between the reviews and interaction data, we propose to align user, item and review-based representations within a shared latent space. As discussed in Sect.~\ref{sec:approach:review-augmented}, a pre-trained language model is used to initialize the review embeddings, which positions the review-based representations in a distinct space from that of  the user/item representations. To reconcile the semantic signals from the reviews and the collaborative signals from the interaction data, it becomes crucial to align the tripartite representations of users, items and reviews in a unified space. Such alignment facilitates a more synergistic integration, so that the semantic signals can effectively complement the collaborative signals.  

Our alignment model boils down to another self-supervised contrastive learning module, inspired by prior multi-modal learning in other fields \cite{G2P2, CLIP}. On one hand, we obtain the user and item representations, such as $\mathbf{e}_u$ for user $u$ and $\mathbf{e}_i$ for item $i$, from the collaborative backbone. On the other hand, the view-level representations derived from the reviews, such as  $\mathbf{h}^k_{u}$ for user $u$ and $\mathbf{h}^k_{i}$ for item $i$,  provide an alternative representation for a user or item, since the reviews written by a user, or written for an item, encapsulate the underlying preferences of the user or inherent characteristics of the item. 
Nevertheless, whether the representations are derived from the interaction data or reviews, they fundamentally describe the same user or item. This correspondence motivates us to seek an alignment between these representations, aiming to maximize the agreement between the representations of the same user or item, while minimizing that between the representations of different users or items. Specifically, we optimize the following losses. 
\begin{align}
L_\text{align}^\text{user} &= \sum_{u\in U}  - \log \frac{\exp(\operatorname{sim}(\mathbf{e}_{u},\mathbf{h}_{u})/\tau)}{ \sum_{u'\in U}\exp(\operatorname{sim}(\mathbf{e}_{u},\mathbf{h}_{u'}))/\tau)) }, \label{eq:align_user}\\
L_\text{align}^\text{item} &= \sum_{i\in I}  - \log \frac{\exp(\operatorname{sim}(\mathbf{e}_{i},\mathbf{h}_{i})/\tau)}{ \sum_{i'\in I}\exp(\operatorname{sim}(\mathbf{e}_{i},\mathbf{h}_{i'}))/\tau)) }, \label{eq:align_item}
\end{align}
where $\mathbf{h}_u$ denotes the representation of user $u$ derived from the reviews. In our implementation, we randomly sample a view from the reviews of user $u$ and employ the view's representation, i.e., $\mathbf{h}_u=\mathbf{h}^1_u$ or $\mathbf{h}^2_u$. Similarly, $\mathbf{h}_i=\mathbf{h}^1_i$ or $\mathbf{h}^2_i$.

\textsc{Remark on missing reviews.} Our review-augmented contrastive strategy is more robust to missing reviews than direct utilization of reviews as input features. Employing reviews as explicit user or item input features (e.g., bag-of-words or dense embeddings) implies that users or items without reviews must resort to dummy or imputed feature values, or rely solely on collaborative embeddings, which may compromise the robustness of learning. 

On the other hand, in our contrastive formulation, if a user or item has no review, they are excluded from the contrastive losses, i.e., skipped in the summation terms in Eqs.~\eqref{eq:rev_user}--\eqref{eq:align_item}, without requiring substitution with dummy or imputed values. 
Note that, despite the exclusion, users or items without any review can still implicitly benefit from the reviews of others, as the entire user-item space is aligned with the review-based representations.

\subsection{Training and Inference} \label{sec:approach:overallloss}

\stitle{Training.}
Finally, we jointly optimize the supervised collaborative loss and the two contrastive losses from both user and item sides.
Specifically, we integrate these losses into an overall loss:
\begin{align}
L = L_\text{CF} + \lambda_1 \left(L_\text{rev}^\text{user}+L_\text{rev}^\text{item}\right) + \lambda_2 \left(L_\text{align}^\text{user}+L_\text{align}^\text{item}\right) + \lambda_3 \|\Theta\|_2^2,\label{eq:loss_total}
\end{align}
where $\Theta$ represent the set of all model parameters, and $\lambda_1,\lambda_2, \lambda_3$ are hyperparameters to control the importance of the review-augmented contrastive learning, alignment contrastive learning, and $L_2$ regularization, respectively.

We outline the pseudocode of the training process and its complexity analysis in Appendix B. 

\stitle{Inference.}
We follow BPR \cite{bpr}, a ranking-based collaborative filtering approach. Specifically, we employ user and item vectors, $\mathbf{e}_u$ and $\mathbf{e}_i$, to predict a ranking score based on their inner product, $\hat{y}_{u,i}=\mathbf{e}_u^\top \mathbf{e}_i$. Note that we employ $\mathbf{e}_*$ alone to capture the primary user/item collaborative signals. Although $\mathbf{h}_*$, the review-based embeddings, are not directly used for ranking, they implicitly improve the collaborative embeddings $\mathbf{e}_*$ through the alignment loss. Nevertheless, a combination of both $\mathbf{e}_*$ and $\mathbf{h}_*$ is also possible when all users and items have reviews. Empirically, using the primary modality for the final prediction can yield good performance.  

%% file: container/_results.tex
\begin{table}[t]
\small
\caption{Statistics of the datasets.}
\vspace{-2mm}
\begin{tabular}{c|r|r|r|r|r}
\hline
Dataset & \#User & \#Items & \# Interactions & \# Reviews & Density   \\ \hline
Kindle                 & 365,687 & 356,888 &3,348,523 &  3,327,676  & .003\% \\ 
Book                   & 500,000 & 678,705 &11,417,323 & 11,266,408  & .003\% \\ 
Beauty                 & 6,119  & 5,082  &23,291 &   22,257  & .075\% \\
Yelp                 & 22,450  & 17,213  &606,123 &   601,512  & .157\% \\ \hline
\end{tabular}%
\label{tab:dataset}
\end{table}
\begin{table*}[t]
\centering
\caption{Performance comparison between ReCAFR and baselines that do not use reviews. Each baseline X is compared with ReCAFR+X, with the better result in each pair bolded. }
\vspace{-2mm}
\small
\addtolength{\tabcolsep}{.5mm}
\begin{tabular}{r|cc|cc|cc|cc}
\hline \hline
  \multicolumn{1}{l|}{\textbf{}} &
  \multicolumn{2}{c|}{\textbf{Kindle}} &
  \multicolumn{2}{c|}{\textbf{Book}} &
  \multicolumn{2}{c|}{\textbf{Beauty}} &
  \multicolumn{2}{c}{\textbf{Yelp}} \\ \hline 
  \textbf{Methods} &
  Recall@5 &
  NDCG@5 &
  Recall@5 &
  NDCG@5 &
  Recall@5 &
  NDCG@5 &
  Recall@5 &
  NDCG@5 \\ \hline
\rowcolor{verylightgray}
  BPR &
  .0203±.0003 &
  .0242±.0005 &
  .0142±.0009 &
  .0236±.0001 &
  .0264±.0007 &
  .0310±.0003 &
  .0109±.0006 &
  .0166±.0002 \\
 \rowcolor{verylightgray}
  ReCAFR+BPR &
  \textbf{.0226}±.0001 &
  \textbf{.0245}±.0001 &
  \textbf{.0143}±.0002 &
  \textbf{.0237}±.0008 &
  \textbf{.0277}±.0003 &
  \textbf{.0313}±.0008 &
  \textbf{.0115}±.0008 &
  \textbf{.0169}±.0004 \\
  LightGCN &
  .0232±.0001 &
  .0281±.0001 &
  .0152±.0005 &
  .0254±.0006 &
  .0254±.0008 &
  .0298±.0000 &
  .0112±.0006 &
  .0168±.0002 \\
  ReCAFR+LightGCN &
  \textbf{.0243}±.0005 &
  \textbf{.0293}±.0001 &
  \textbf{.0187}±.0005 &
  \textbf{.0311}±.0001 &
  \textbf{.0266}±.0006 &
  \textbf{.0300}±.0006 &
  \textbf{.0125}±.0006 &
  \textbf{.0175}±.0005 \\
 \rowcolor{verylightgray}
  SGL &
  .0237±.0006 &
  .0286±.0007 &
  .0169±.0001 &
  .0281±.0005 &
  .0269±.0002 &
  .0316±.0008 &
  .0101±.0004 &
  .0151±.0004 \\
 \rowcolor{verylightgray}
  ReCAFR+SGL &
  \textbf{.0242}±.0009 &
  \textbf{.0291}±.0005 &
  \textbf{.0171}±.0006 &
  \textbf{.0285}±.0007 &
  \textbf{.0276}±.0007 &
  \textbf{.0319}±.0009 &
  \textbf{.0106}±.0001 &
  \textbf{.0158}±.0001 \\ 
  DirectAU &
  .0255±.0009 &
  .0309±.0007 &
  .0167±.0009 &
  .0281±.0009 &
  .0298±.0007 &
  .0338±.0006 &
  \textbf{.0124}±.0009 &
  .0169±.0005 \\
  ReCAFR+DirectAU &
  \textbf{.0262}±.0000 &
  \textbf{.0317}±.0009 &
  \textbf{.0172}±.0006 &
  \textbf{.0285}±.0007 &
  \textbf{.0301}±.0008 &
  \textbf{.0341}±.0007 &
  .0121±.0006 &
  \textbf{.0172}±.0008 \\
   \rowcolor{verylightgray}
  SimGCL &
  .0253±.0002 &
  \textbf{.0306}±.0003 &
  .0180±.0007 &
  .0301±.0007 &
  .0288±.0000 &
  .0339±.0007 &
  .0145±.0001 &
  .0188±.0002 \\
 \rowcolor{verylightgray}
  ReCAFR+SimGCL &
  \textbf{.0269}±.0002 &
  .0302±.0003 &
  \textbf{.0184}±.0003 &
  \textbf{.0307}±.0001 &
  \textbf{.0296}±.0007 &
  \textbf{.0365}±.0001 &
  \textbf{.0155}±.0007 &
  \textbf{.0202}±.0004 \\  
  \hline \hline
\textbf{Methods} &
  Recall@20 &
  NDCG@20 &
  Recall@20 &
  NDCG@20 &
  Recall@20 &
  NDCG@20 &
  Recall@20 &
  NDCG@20 \\ \hline 
  BPR &
  \textbf{.0571}±.0001 &
  .0394±.0001 &
  .0439±.0009 &
  .0340±.0001 &
  .0750±.0009 &
  .0516±.0005 &
  .0299±.0007 &
  .0234±.0006 \\
  ReCAFR+BPR &
  .0553±.0009 &
  \textbf{.0399}±.0009 &
  \textbf{.0440}±.0004 &
  \textbf{.0351}±.0006 &
  \textbf{.0821}±.0003 &
  \textbf{.0542}±.0002 &
  \textbf{.0315}±.0002 &
  \textbf{.0241}±.0007 \\
 \rowcolor{verylightgray}
  LightGCN &
  .0616±.0002 &
  .0437±.0007 &
  .0472±.0006 &
  .0365±.0006 &
  .0720±.0002 &
  .0496±.0006 &
  .0304±.0001 &
  .0237±.0005 \\
 \rowcolor{verylightgray}
  ReCAFR+LightGCN &
  \textbf{.0645}±.0009 &
  \textbf{.0457}±.0003 &
  \textbf{.0543}±.0002 &
  \textbf{.0430}±.0003 &
  \textbf{.0788}±.0007 &
  \textbf{.0520}±.0004 &
  \textbf{.0316}±.0000 &
  \textbf{.0241}±.0008 \\
  SGL &
  .0628±.0006 &
  .0446±.0001 &
  .0523±.0004 &
  .0404±.0001 &
  \textbf{.0763}±.0002 &
  \textbf{.0525}±.0009 &
  .0273±.0004 &
  \textbf{.0213}±.0006 \\
  ReCAFR+SGL &
  \textbf{.0629}±.0006 &
  \textbf{.0448}±.0003 &
  \textbf{.0525}±.0009 &
  \textbf{.0412}±.0001 &
  .0761±.0002 &
  .0513±.0008 &
  \textbf{.0278}±.0003 &
  .0211±.0005 \\ 
  \rowcolor{verylightgray}
  DirectAU &
  \textbf{.0681}±.0007 &
  .0481±.0007 &
  \textbf{.0529}±.0005 &
  .0411±.0007 &
  .0793±.0007 &
  .0546±.0006 &
  \textbf{.0307}±.0003 &
  \textbf{.0238}±.0007 \\
 \rowcolor{verylightgray}
  ReCAFR+DirectAU &
  .0676±.0004 &
  \textbf{.0491}±.0002 &
  .0516±.0005 &
  \textbf{.0418}±.0005 &
  \textbf{.0801}±.0005 &
  \textbf{.0549}±.0003 &
  .0302±.0003 &
  .0223±.0001 \\
  SimGCL &
  .0659±.0006 &
  .0468±.0006 &
  .0549±.0006 &
  .0424±.0001 &
  .0801±.0008 &
  .0551±.0004 &
  .0340±.0007 &
  .0266±.0007 \\
  ReCAFR+SimGCL &
  \textbf{.0670}±.0003 &
  \textbf{.0476}±.0009 &
  \textbf{.0582}±.0002 &
  \textbf{.0440}±.0001 &
  \textbf{.0839}±.0001 &
  \textbf{.0564}±.0005 &
  \textbf{.0349}±.0006 &
  \textbf{.0277}±.0009 \\
  \hline \hline
\end{tabular}%
\label{tab:overall}
\end{table*}

\section{Experiments}\label{sec:expt}

We first describe the experimental settings, then evaluate ReCAFR and various baseline methods in detail, followed by additional analyses of our approach.

\subsection{Experimental Setup} \label{sect:expt:setup}
\stitle{Datasets.}
We conduct experiments on four benchmark  datasets in various domains, namely, ``\textbf{Kindle} Store'', ``\textbf{Book}s'', ``All \textbf{Beauty}'' from Amazon \cite{ni-etal-2019-justifying} 
and \textbf{Yelp}\footnote{\url{https://www.yelp.com/dataset}}. 
We adopt the 2-core setting for Kindle, Beauty and Yelp, and 5-core for Book, where
the $K$-core setting ensures that each user and item has at least $K$ interactions. 
We set a smaller $K$ for Kindle and Beauty as they have a smaller average number of interactions per user or item.
For Book, we randomly sample 500K users from the 5-core setting. 
For Yelp, we utilize the Phoenix subset. 
We further process the raw reviews by removing those consisting solely of non-letter characters. Each dataset is randomly divided into training, validation, and test sets, following an 80\%, 10\%, and 10\% split. The statistics of the processed datasets are summarized in Table~\ref{tab:dataset}.

\stitle{Baseline methods.}
We compare to the following baselines in two broad categories, as follows.

The first category involves collaborative filtering  without using reviews. 
(1) \textbf{BPR} \cite{bpr}: A classical matrix factorization technique optimized by the BPR loss. 
(2) \textbf{LightGCN} \cite{4}: A graph-based approach specifically designed to improve recommendation performance on a user-item bipartite graph.
(3) \textbf{SGL} \cite{sgl}: A self-supervised learning approach on the user-item graph, leveraging structural augmentations involving node/edge dropout and random walk. 
(4) \textbf{DirectAU} \cite{DirectAU}: A state-of-the-art approach, optimizing the desirable properties of representations in collaborative filtering on the unit hypersphere.
(5) \textbf{SimGCL} \cite{SimGCL}: A state-of-the-art approach with augmentation-free contrastive learning for recommendation.
Note that we can employ any of these approaches as the collaborative backbone for ReCAFR. 

The second category involves recent review-aware approaches. 
(1) \textbf{NARRE} \cite{NARRE}: An attention-based approach that extracts and aggregates user and item features from reviews. 
(2) \textbf{RGCL} \cite{RGCL}: A state-of-the-art review-aware graph-based contrastive method. However, their contrastive augmentations using node and edge discrimination rely primarily on graph structures, involving reviews merely as features for the node/edge discrimination tasks. 
(3) \textbf{RMCL} \cite{RMCL}: A state-of-the-art intention representation method based on the mixed Gaussian distribution hypothesis, with a multi-intention contrastive strategy.
For these review-aware baselines, 
to deal with users or items without reviews, we impute a dummy feature vector based on the mean pooling of 10K randomly sampled reviews.

\begin{table*}[t]
\centering
\caption{Performance comparison between ReCAFR and review-aware baselines. The best result in each setting is bolded.}
\vspace{-2mm}
\small
\addtolength{\tabcolsep}{0.6mm}
\begin{tabular}{lcc|cc|cc|cc}
\hline \hline
  \multicolumn{1}{c|}{\textbf{Method}} &
  \multicolumn{2}{c|}{\textbf{Kindle}} &
  \multicolumn{2}{c|}{\textbf{Book}} &
  \multicolumn{2}{c|}{\textbf{Beauty}} &
  \multicolumn{2}{c}{\textbf{Yelp}} \\ \hline 
    \multicolumn{9}{c}{\textbf{Using all available reviews}} \\ \hline 
  \multicolumn{1}{l|}{} &
  Recall@5 &
  NDCG@5 &
  Recall@5 &
  NDCG@5 &
  Recall@5 &
  NDCG@5 &
  Recall@5 &
  NDCG@5 \\ \hline
  \multicolumn{1}{l|}{NARRE} &
  .0238±.0009 &
  \multicolumn{1}{c|}{.0269±.0007} &
  .0160±.0008 &
  \multicolumn{1}{c|}{.0266±.0001} &
  .0266±.0006 &
  \multicolumn{1}{c|}{.0303±.0009} &
  .0129±.0002 &
  .0171±.0006 \\
  \multicolumn{1}{l|}{RGCL} &
  .0242±.0006 &
  \multicolumn{1}{c|}{.0292±.0001} &
  .0183±.0003 &
  \multicolumn{1}{c|}{.0305±.0003} &
  .0269±.0006 &
  \multicolumn{1}{c|}{.0325±.0004} &
  .0142±.0002 &
  .0196±.0005 \\
  \multicolumn{1}{l|}{RMCL} &
  .0236±.0008 &
  \multicolumn{1}{c|}{.0288±.0001} &
  .0172±.0003 &
  \multicolumn{1}{c|}{.0294±.0009} &
  .0273±.0006 &
  \multicolumn{1}{c|}{.0315±.0004} &
  .0144±.0007 &
  .0192±.0006 \\
  \multicolumn{1}{l|}{ReCAFR+SimGCL} &
  \textbf{.0269}±.0004 &
  \multicolumn{1}{c|}{\textbf{.0302}±.0003} &
  \textbf{.0184}±.0005 &
  \multicolumn{1}{c|}{\textbf{.0307}±.0003} &
  \textbf{.0296}±.0003 &
  \multicolumn{1}{c|}{\textbf{.0365}±.0008} &
  \textbf{.0155}±.0009 &
  \textbf{.0202}±.0007 \\ \hline
\hline
\multicolumn{9}{c}{\textbf{Removing 30\% of the reviews}} \\ \hline 
  \multicolumn{1}{l|}{} &
  Recall@5 &
  NDCG@5 &
  Recall@5 &
  NDCG@5 &
  Recall@5 &
  NDCG@5 &
  Recall@5 &
  NDCG@5 \\ \hline
  \multicolumn{1}{l|}{NARRE} &
  .0206±.0007 &
  \multicolumn{1}{c|}{.0213±.0008} &
  .0141±.0005 &
  \multicolumn{1}{c|}{.0211±.0006} &
  .0239±.0007 &
  \multicolumn{1}{c|}{.0240±.0006} &
  .0102±.0006 &
  .0140±.0007 \\
  \multicolumn{1}{l|}{RGCL} &
  .0215±.0009 &
  \multicolumn{1}{c|}{.0216±.0009} &
  .0152±.0002 &
  \multicolumn{1}{c|}{.0273±.0006} &
  .0247±.0005 &
  \multicolumn{1}{c|}{.0248±.0008} &
  .0109±.0001 &
  .0156±.0001 \\
  \multicolumn{1}{l|}{RMCL} &
  .0226±.0007 &
  \multicolumn{1}{c|}{.0211±.0005} &
  .0163±.0005 &
  \multicolumn{1}{c|}{.0259±.0001} &
  .0251±.0001 &
  \multicolumn{1}{c|}{.0264±.0008} &
  .0116±.0005 &
  .0148±.0005 \\
  \multicolumn{1}{l|}{ReCAFR+SimGCL} &
  \textbf{.0241}±.0007 &
  \multicolumn{1}{c|}{\textbf{.0279}±.0007} &
  \textbf{.0171}±.0001 &
  \multicolumn{1}{c|}{\textbf{.0295}±.0004} &
  \textbf{.0274}±.0008 &
  \multicolumn{1}{c|}{\textbf{.0329}±.0006} &
  \textbf{.0121}±.0005 &
  \textbf{.0184}±.0001 \\ \hline
\hline
\end{tabular}%
\label{tab:review-aware}
\end{table*}

\stitle{Evaluation.}
We evaluate the ranking performance of top-$K$ recommendations with Recall$@K$, Precision$@K$, and NDCG$@K$, for $K\in\{5,20\}$. For each metric, we compute the average over all users in the test set. Furthermore, each experiment is repeated five times and the means and standard deviations of the performance over the five runs are reported.
For a fair comparison, we have adapted all methods to optimize the same BPR loss as used in our approach, which is appropriate for the ranking-based evaluation. Notably, our experiments show that the adapted versions generally outperform their original loss functions.

\stitle{Implementation details.} 
ReCAFR can flexibly adopt any supervised collaborative backbone. 
Given a backbone X, we denote our ReCAFR paired with the backbone as ReCAFR+X. Specifically, 
we adopt each of the five baselines that do not use reviews (BPR, LightGCN, SGL, SimGCL, DirectAU) as our backbone.

For ReCAFR, the embedding dimension is set as $d=64$. To sample two views from the reviews of a user or item, we randomly split the reviews into two subsets of equal size, treating each subset as one view. To initialize the embedding of each review, we use a pre-trained sentence-transformers model \cite{reimers-2019-sentence-bert},
denoted as $f(\cdot)$ in Eqs.~\eqref{eq:view_embedding_u}--\eqref{eq:view_embedding_i}.
We also use the average embedding vectors of the reviews associated with a user or item  to initialize their embeddings for the collaborative backbone. For users or items without any review, we initialize their embeddings with a dummy vector. We select the hyperparameters through a grid search over $\lambda_1,\lambda_2\in\{.001, .01, 1, 10, 20\}$ and $\lambda_3\in\{0.1,0.2,0.3,\ldots,1.0\}$ on the validation set.
The temperature $\tau$ in the contrastive loss is searched from $0$ to $1$.
We employ Adam \cite{adam} as the optimizer for the training process with the batch size fixed at 1024 and the learning rate at $.01$.  

For the various baselines, we defer their implementation details to Appendix~C. 


\subsection{Performance Comparison}

We compare ReCAFR to the baselines without or with reviews. Due to space limitations, only selected metrics are shown, with the full results provided in Appendix E.

\stitle{Comparison to baselines w/o reviews}. Table~\ref{tab:overall} presents the results of ReCAFR and five collaborative baselines that do not utilize any reviews. We pair each baseline with ReCAFR as its backbone,
and evaluate the performance of each pair. 

We generally observe better performance when each backbone recommender is integrated with ReCAFR than when it is not, demonstrating the effectiveness of ReCAFR.
Furthermore, contrastive backbones such as SGL and SimGCL generally outperform conventional collaborative filtering models like BPR and LightGCN. 
This is because contrastive methods leverage self-supervised signals to learn inherent properties within the interaction data, which complement supervised collaborative signals. 
Hence, by incorporating additional self-supervised signals from the reviews, ReCARF could achieve even better performance.
In particular, when paired with the state-of-the-art SimGCL, ReCARF+SimGCL generally outperform all other variants in our experiments. 

\begin{figure}[t]
\centering
\includegraphics[width=0.99\columnwidth]{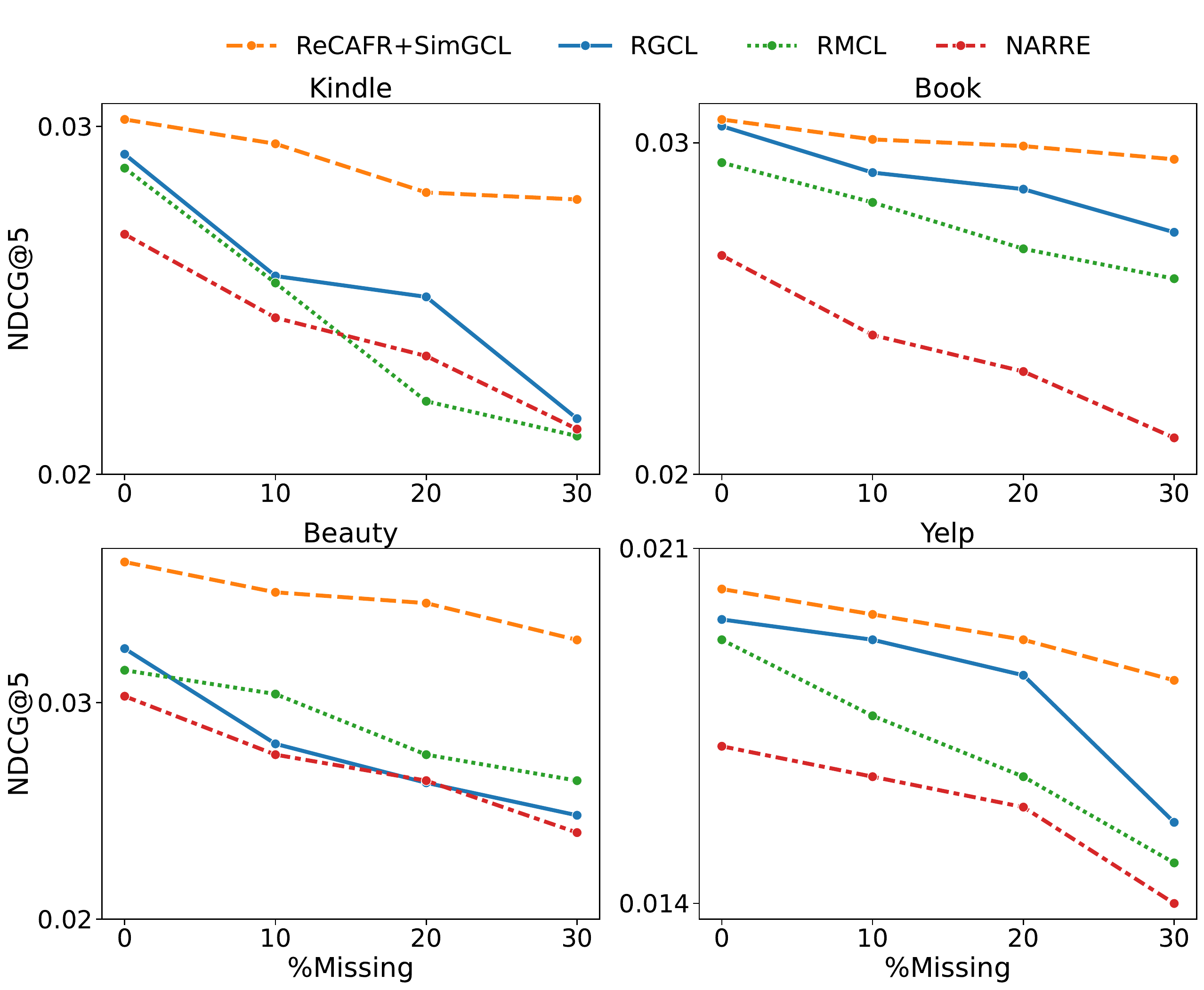}
\vspace{-2mm}
\caption{Impact of missing reviews.}
\label{fig:missing review}
\end{figure}

\stitle{Comparison to review-aware baselines.} Next, we compare ReCARF+SimGCL, our most competitive variant, with various review-aware methods. 
In Table~\ref{tab:review-aware}, we include two settings: using all available reviews, and removing 30\% of the reviews. The rationale for removing the reviews is to simulate real-world scenarios where many interactions lack reviews, noting that the public benchmarks were curated in a way that only interactions with reviews were harvested, which does not represent real-world scenarios.  

First, we consistently observe that ReCAFR demonstrates superior performance in all cases. The advantage  of ReCAFR suggests that integrating reviews with collaborative filtering within a unified space can more effectively mitigate the sparsity of interaction data, compared to other review-aware methods that merely extract features from reviews. 

Second, when 30\% of the reviews are removed, all methods experience a performance drop; however, ReCAFR tends to be more robust, with smaller decreases. This further demonstrates the robustness of our approach when facing missing reviews. To more comprehensively evaluate the robustness in the absence of reviews, we vary the amount of reviews removed from each dataset, and visualize the performance trend in Fig.~\ref{fig:missing review}. 
We observe that, as more reviews are removed, the performance of all review-aware methods declines. Notably, NARRE, RGCL, and RMCL exhibit a more pronounced decrease in performance, whereas our method, ReCAFR, experiences a significantly smaller reduction, reaffirming its robustness in dealing with missing reviews.

\subsection{Ablation Study}

We conduct an ablation study on ReCAFR with the SimGCL backbone, noting that using other backbones shows similar patterns. To evaluate the contribution of our key designs in ReCAFR, we compare with the following ablated variants.
(1) \textbf{w/o text emb.~init.}: We do not use pre-trained review embeddings to initialize user and item representations in the collaborative backbone. Instead, we randomly initialize them.
(2) \textbf{w/o user CL}: We eliminate the user-side contrastive learning, including both the review-augmented and alignment contrastive strategies, while retaining the item-side contrastive learning. Note that removing either contrastive strategy in isolation is not feasible; without review augmentation, alignment becomes redundant as review-based representations are absent. Conversely, omitting the alignment strategy means that user and item representations are optimized on the interaction data independently, disregarding review augmentation. 
(Nevertheless, we demonstrate the relative contribution from each strategy in Sect.~\ref{sec:expt:hyperparam}.) 
(3) \textbf{w/o item CL}: We remove the item-side contrastive learning, while retaining the user-side contrastive learning.

We present the performance of the full model and these variants in Table~\ref{tab:ablation}, and make the following observations. First, 
we observe that the variant ``w/o text emb.~init.'' shows a consistent decrease in performance compared to the full model. This outcome is expected as, pre-trained review-based embeddings contain useful semantic information that helps bootstrap model training. However, relatively modest performance decrease suggests that using reviews merely as side information does not fully tap into the potential of reviews. 
Second, the results reveal that the variants without either user- or item-side contrastive learning suffer from a notable decline in performance compared to the full model. This outcome substantiates our thesis that the proposed contrastive strategies can effectively complement the interaction data. 


\begin{table}[t]
\centering
\small
\caption{Ablation study on ReCAFR, reporting NDCG@5.}
\vspace{-2mm}
\addtolength{\tabcolsep}{1mm}
\begin{tabular}{l|c|c|c|c}
\hline 
\textbf{Variants}   & \textbf{Kindle} & \textbf{Book}   & \textbf{Beauty} & \textbf{Yelp}   \\ \hline 
ReCAFR+SimGCL     & \textbf{.0302}          & \textbf{.0307} & \textbf{.0365} & \textbf{.0202} \\
w/o text emb. init. & .0294          & .0286          & .0351          & .0199          \\
w/o user CL         & .0281          & .0281          & .0346          & .0186          \\
w/o item CL         & .0276          & .0284          & .0331          & .0182          \\\hline
\end{tabular}%
\label{tab:ablation}
\end{table}





\begin{figure}[t]
\centering
\includegraphics[width=0.99\columnwidth]{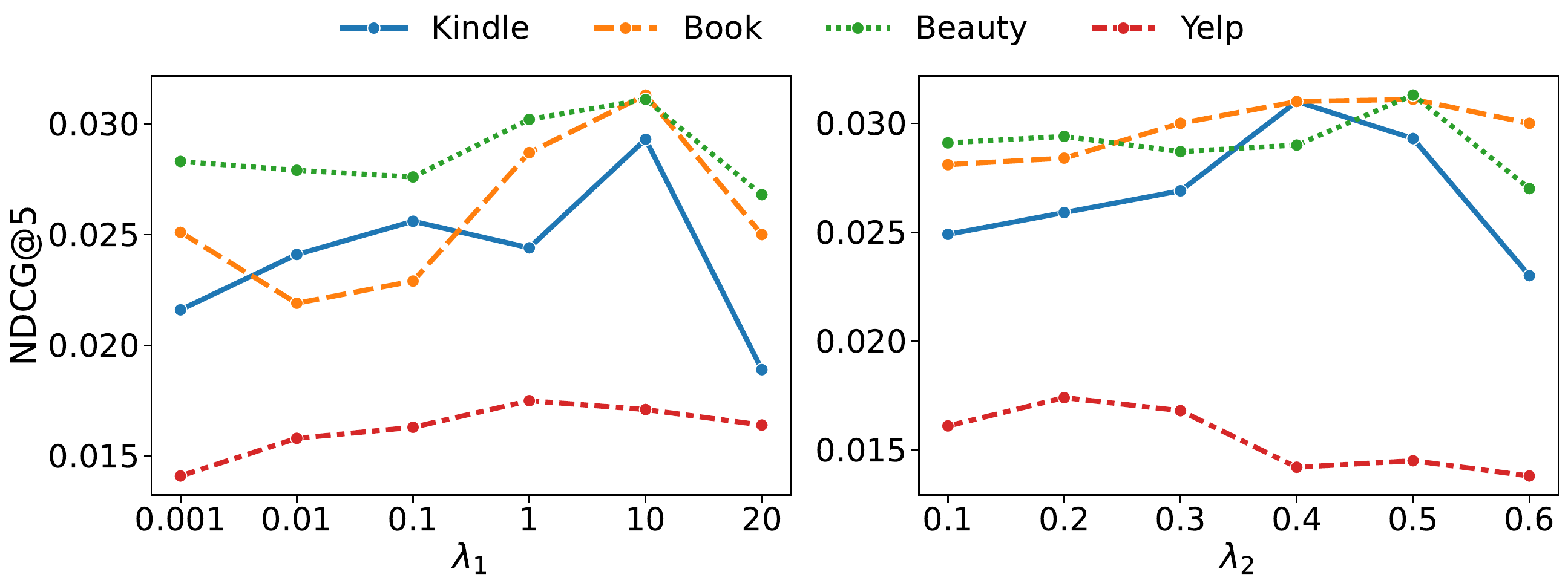}
\vspace{-2mm}
\caption{Effect of $\lambda_1$ and $\lambda_2$ on ReCAFR.}
\label{fig:lambda1}
\end{figure}


\subsection{Hyperparameter Studies}\label{sec:expt:hyperparam}
 
We investigate the impact of $\lambda_1$ and $\lambda_2$ in Eq.~\eqref{eq:loss_total}, the two main hyperparameters in ReCAFR. 
Given the similarity in results from various backbones, we only report the findings obtained with the LightGCN backbone in Fig.~\ref{fig:lambda1}.

First, to analyze the influence of $\lambda_1$, we vary it in the range of $.001$ to $20$, while fixing $\lambda_2 = 0.5$ across all datasets. As shown in Fig.~\ref{fig:lambda1} (left), an appropriate $\lambda_1$ value can effectively improve the performance of ReCAFR. Specifically, ReCAFR demonstrates strong performance across four datasets when $\lambda_1$ is set around 10. This observation implies that our review-augmented contrastive strategy plays a pivotal role,
and underscores the need of a balanced trade-off between the supervisory signals in review data and the semantic signals in interaction data.

Next, we fix $\lambda_1 = 10$, and vary $\lambda_2$ in the range of $0.1$ to $0.6$. ReCAFR appears stable when $\lambda_2$ is set around the range  $[0.4,0.5]$ on the Amazon datasets and $0.2$ on Yelp. This outcome suggests that our alignment contrastive strategy is a crucial factor in the learning process: insufficient alignment (i.e., smaller $\lambda_2$) leads to inconsistent latent spaces, while excessive alignment (i.e., larger $\lambda_2$) imposes too much constraint on the optimization, thereby diminishing the model's capacity.

\begin{table}[]
\centering
\small
\caption{Demonstration of ReCAFR with LLM-enhanced reviews on the Beauty dataset.}
\vspace{-2mm}
\addtolength{\tabcolsep}{1mm}
\begin{tabular}{l|ccc}
\hline
Methods                & Recall@5             & Prec@5             & NDCG@5          \\ \hline
BPR                      & .0264          & .0272          & .0310           \\
ReCAFR+BPR             & \textbf{.0277} & .0287          & .0313          \\
ReCAFR+BPR (LLM)       & .0269          & \textbf{.0289} & \textbf{.0315} \\ \hline
LightGCN                & .0254          & .0261          & .0298          \\
ReCAFR+LightGCN       & .0266          & .0276          & .0300            \\
ReCAFR+LightGCN (LLM) & \textbf{.0269} & \textbf{.0277} & \textbf{.0311} \\ \hline
DirectAU                & .0298          & .0284          & .0338          \\
ReCAFR+DirectAU       & .0301          & .0286          & .0341          \\
ReCAFR+DirectAU (LLM) & \textbf{.0306} & \textbf{.0295} & \textbf{.0349} \\ \hline
SimGCL                  & .0288          & .0295          & .0338          \\
ReCAFR+SimGCL         & .0296          & .0304          & .0365          \\
ReCAFR+SimGCL (LLM)   & \textbf{.0298} & \textbf{.0313} & \textbf{.0376} \\ \hline
\end{tabular}%
\label{tab:LLMs}
\end{table}

\subsection{Can ReCAFR benefit from LLMs?}\label{sect:expt:llm}

With the emergence of large language models (LLMs), an orthogonal direction is whether LLMs could be used to enhance reviews.  In this final part, we conduct a preliminary analysis on whether ReCAFR can further benefit from LLM-enhanced reviews. LLMs can summarize key points, remove noises, and perform reasoning on the reviews, potentially improving their overall quality.
To this end, we follow a previous work called RLMRec \cite{RLMRec}, 
using well-designed LLM instructions and input prompts to refine and enhance the reviews. Specifically, we query an LLM (GPT-4o mini) using such instructions and prompts, generating a response that includes both ``summarization'' and  ``reasoning'' sections for each review. 
The generated responses represent a refined version of the reviews.
Details of the instruction and prompt are described in Appendix D.  

In Table~\ref{tab:LLMs}, we present the results using several collaborative backbones on the Beauty dataset.
``ReCAFR+X (LLM)'' indicates the use of LLM-enhanced reviews in place of the original reviews as input.
Overall, LLM-enhanced reviews consistently boost ReCAFR's performance across various backbones, suggesting the feasibility of improving recommendations with LLMs, as well as demonstrating the flexibility and robustness of ReCAFR in different settings. Since the LLM-enhancement is an orthogonal notion that is not integral to the algorithmic method itself, we leave further investigation on the integration of LLMs for recommendation to future work.


%% file: container/_conclusion.tex
\section{Conclusion and Future Work}
In this work, we proposed a Review-centric Contrastive Alignment Framework for Recommendation (ReCAFR). Distinct from existing  approaches that largely treat reviews as input features, ReCAFR incorporates reviews into collaborative filtering within a contrastive framework to enhance the synergy between reviews and interaction data. Specifically, we introduced two self-supervised contrastive strategies that not only utilize the review augmentation to alleviate sparsity but also align the tripartite representations within a unified space to better exploit the interplay among users, items and reviews. In future work, we plan to extend our framework to incorporate other contrastive signals, such as graph-based self-supervision. 

%% file: container/_appendix.tex
\section{Details of the pilot study} \label{app:pilot}
We describe the details of our pilot study in Fig.~1, which shows the contrast between the similarity distributions of positive pairs and negative pairs. 

As elaborated in experimental setup (see Sect.~5.1), we first sample two views from the set of reviews of each user $u$, denoted by $S^1_u$ and $S^2_u$. Then, we employ a pre-trained sentence-transformers model (MiniLM) $f(\cdot)$ to map $S^1_u$ and $S^2_u$ into view-level representations, $\vec{h}^1_u$ and $\vec{h}^2_u$, respectively (Eq.~3). Furthermore, we construct positive and negative pairs from the views of all users following our proposed review augmentation process (see Sect.~4.2). Given the view-level representations and the augmented pairs, we compute the cosine similarity of the positive pairs, e.g., $\text{sim}(\vec{h}^1_u,\vec{h}^2_u)$, and that of the negative pairs, e.g., $\text{sim}(\vec{h}^1_u,\vec{h}^2_{u'})$. Finally, we plot the similarity distribution curves of the positive pairs, and that of the negative pairs, on each dataset. We also obtain similar patterns with different pre-trained language models including BERT and SBERT.

\section{Pseudocode} \label{app:pseudocode}

The pseudocode of the training process is outlined in Algorithm~\ref{alg:ReCAFR}.
The overall complexity can be analyzed based on the computation of  three major components: the collaborative backbone, the review-augmented contrastive learning, and the alignment contrastive learning.
The complexity of the contrastive backbone grows linearly with the training set, i.e., $O(|\mathcal{O}|)$.
For the two contrastive losses, if their normalization is computed in full over all the users $u'\in U$ or items $i' \in I$ as shown in Eqs.~(5)--(8), their theoretical complexity is $O(|U|^2+|I|^2)$. However, in practice we perform sampling only (Algorithm~\ref{alg:ReCAFR}, line 9) within each mini-batch, reducing the complexity to $O(|U|+|I|)$.
Hence, the overall complexity of ReCAFR is $O(|\mathcal{O}|+|U|+|I|)$. Comparing to the collaborative backbone, we incur an overhead of $O(|U|+|I|)$ only, which is linear to the number of users and items. Moreover, in general, the number of users and items are much less than the total number of training triples, since each user or item contributes multiple training triples. Hence, the overhead in ReCAFR over the collaborative backbone is negligible, given that $|U|+|I|\ll |\mathcal{O}|$.

\begin{algorithm}[t]
\small
\caption{\textsc{Training Procedure of ReCAFR}}
\label{alg:ReCAFR}
\begin{algorithmic}[1]
   \Require The set of training triplets $\mathcal{O}$;
   \Require Reviews of each user $\{S_u:u\in U\}$ and each item $\{S_i:i \in I\}$.
   \For{each user $u \in U$ (or each item $i \in I$)}
        \State Sample two views $S^1_u,S^2_u$ from $S_u$ (or $S^1_i,S^2_i$ from $S_i$);
        \State Initialize $\textbf{h}^1_u,\textbf{h}^2_u$ (or $\vec{h}^1_i$ and $\vec{h}^2_i$) for the sampled views;
   \EndFor
   \State $\Theta \gets$ model initialization;
   \While{not converged}
        \For{each training triple $(u,i,j)\in \mathcal{O}$} 
            \State Compute $L_\text{CF}$; (Eq.~2)
            \State Sample $u'\in U, i'\in I$;
            \State Compute $L^\text{user}_\text{rev}$ and $L^\text{item}_\text{rev}$ (Eqs.~5, 6);
            \State Compute $L^\text{user}_\text{align}$ and $L^\text{item}_\text{align}$  (Eqs.~7, 8);    
            \State Compute the total loss $L$ (Eq.~9);
        \EndFor
        \State Update $\Theta \gets$ via backpropagation;
    \EndWhile
    \State \Return $\Theta$.
   
\end{algorithmic}
\end{algorithm}

\begin{figure*}[t]
\centering
\includegraphics[width=0.85\textwidth]{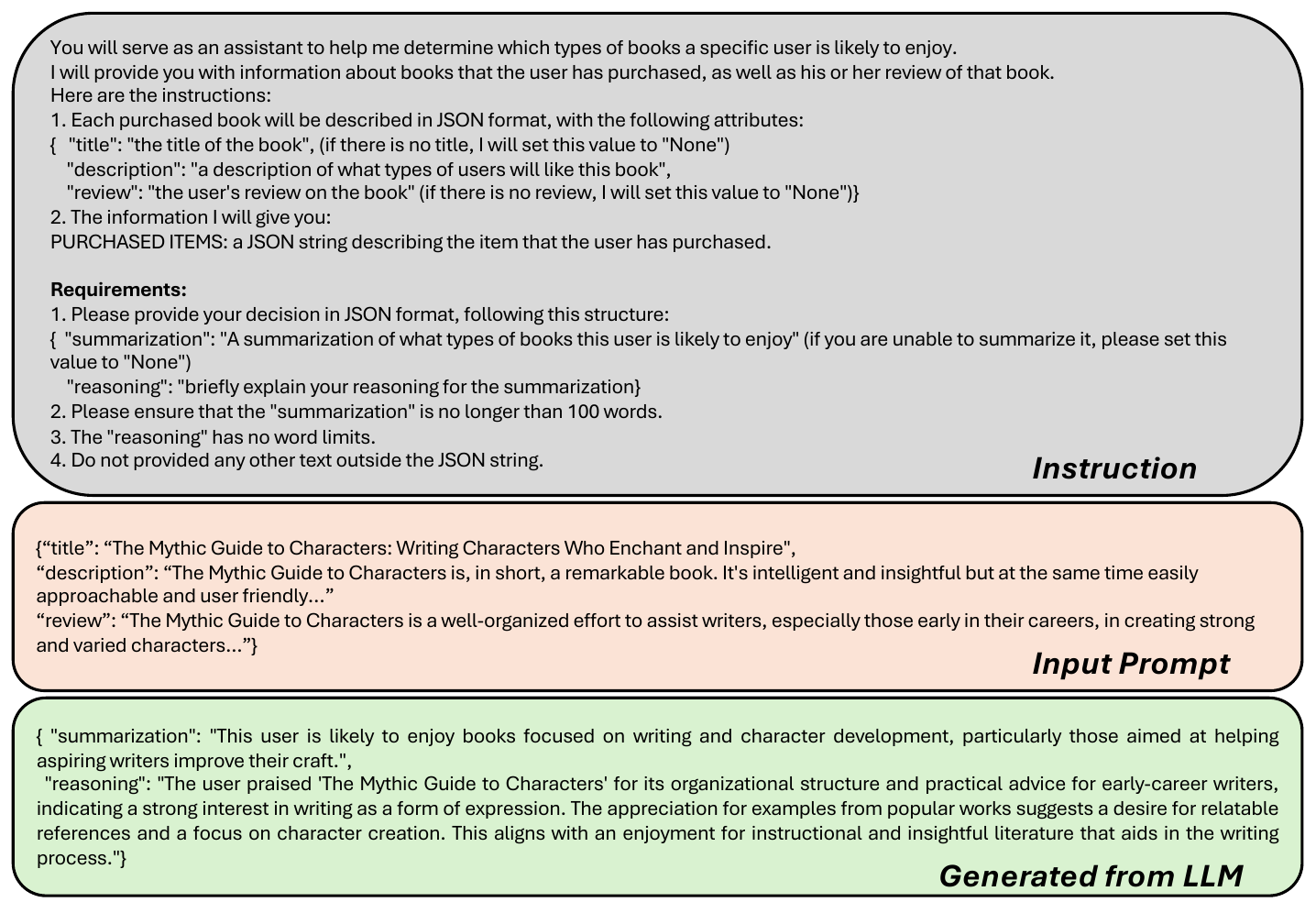}
\vspace{-4mm}
\caption{Review enhancement using LLMs, illustrated with the Book dataset.}
\label{fig:LLMex}
\end{figure*}

\begin{table*}[t]
\centering
\caption{Performance comparison (@5) between ReCAFR and review-based baselines. The best result in each setting is bolded.}
\vspace{-2mm}
\addtolength{\tabcolsep}{-1mm}
\renewcommand{\arraystretch}{1.2}
\resizebox{1\textwidth}{!}{%
\begin{tabular}{lccc|ccc|ccc|ccc}
\hline \hline
  \multicolumn{1}{c|}{\textbf{Method}} &
  \multicolumn{3}{c|}{\textbf{Kindle}} &
  \multicolumn{3}{c|}{\textbf{Book}} &
  \multicolumn{3}{c|}{\textbf{Beauty}} &
  \multicolumn{3}{c}{\textbf{Yelp}} \\ \hline 
    \multicolumn{13}{c}{\textbf{Using all available reviews}} \\ \hline 
  \multicolumn{1}{l|}{} &
  Recall@5 &
  Prec@5 &
  NDCG@5 &
  Recall@5 &
  Prec@5 &
  NDCG@5 &
  Recall@5 &
  Prec@5 &
  NDCG@5 &
  Recall@5 &
  Prec@5 &
  NDCG@5 \\ \hline
  \multicolumn{1}{l|}{NARRE} &
  .0238±.0009 &
  .0235±.0005 &
  \multicolumn{1}{c|}{.0269±.0007} &
  .0160±.0008 &
  .0257±.0003 &
  \multicolumn{1}{c|}{.0266±.0001} &
  .0266±.0006 &
  .0276±.0002 &
  \multicolumn{1}{c|}{.0303±.0009} &
  .0129±.0002 &
  .0182±.0009 &
  .0171±.0006 \\
  \multicolumn{1}{l|}{RGCL} &
  .0242±.0006 &
  .0264±.0009 &
  \multicolumn{1}{c|}{.0292±.0001} &
  .0183±.0003 &
  .0293±.0003 &
  \multicolumn{1}{c|}{.0305±.0003} &
  .0269±.0006 &
  .0279±.0002 &
  \multicolumn{1}{c|}{.0325±.0004} &
  .0142±.0002 &
  .0189±.0006 &
  .0196±.0005 \\
  \multicolumn{1}{l|}{RMCL} &
  .0236±.0008 &
  .0259±.0006 &
  \multicolumn{1}{c|}{.0288±.0001} &
  .0172±.0003 &
  .0286±.0001 &
  \multicolumn{1}{c|}{.0294±.0009} &
  .0273±.0006 &
  .0265±.0006 &
  \multicolumn{1}{c|}{.0315±.0004} &
  .0144±.0007 &
  .0186±.0003 &
  .0192±.0006 \\
  \multicolumn{1}{l|}{ReCAFR+SimGCL} &
  \textbf{.0269}±.0004 &
  \textbf{.0285}±.0008 &
  \multicolumn{1}{c|}{\textbf{.0302±.0003}} &
  \textbf{.0184}±.0005 &
  \textbf{.0295}±.0005 &
  \multicolumn{1}{c|}{\textbf{.0307±.0003}} &
  \textbf{.0296}±.0003 &
  \textbf{.0304}±.0001 &
  \multicolumn{1}{c|}{\textbf{.0365}±.0008} &
  \textbf{.0155}±.0009 &
  \textbf{.0191}±.0008 &
  \textbf{.0202}±.0007 \\ \hline
\hline
\multicolumn{13}{c}{\textbf{Removing 30\% of the reviews}} \\ \hline 
  \multicolumn{1}{l|}{} &
  Recall@5 &
  Prec@5 &
  NDCG@5 &
  Recall@5 &
  Prec@5 &
  NDCG@5 &
  Recall@5 &
  Prec@5 &
  NDCG@5 &
  Recall@5 &
  Prec@5 &
  NDCG@5 \\ \hline
  \multicolumn{1}{l|}{NARRE} &
  .0206±.0007 &
  .0229±.0009 &
  \multicolumn{1}{c|}{.0213±.0008} &
  .0141±.0005 &
  .0214±.0008 &
  \multicolumn{1}{c|}{.0211±.0006} &
  .0239±.0007 &
  .0261±.0007 &
  \multicolumn{1}{c|}{.0240±.0006} &
  .0102±.0006 &
  .0145±.0005 &
  .0140±.0007 \\
  \multicolumn{1}{l|}{RGCL} &
  .0215±.0009 &
  .0238±.0004 &
  \multicolumn{1}{c|}{.0216±.0009} &
  .0152±.0002 &
  .0258±.0009 &
  \multicolumn{1}{c|}{.0273±.0006} &
  .0247±.0005 &
  .0274±.0008 &
  \multicolumn{1}{c|}{.0248±.0008} &
  .0109±.0001 &
  .0152±.0005 &
  .0156±.0001 \\
  \multicolumn{1}{l|}{RMCL} &
  .0226±.0007 &
  .0248±.0002 &
  \multicolumn{1}{c|}{.0211±.0005} &
  .0163±.0005 &
  .0247±.0003 &
  \multicolumn{1}{c|}{.0259±.0001} &
  .0251±.0001 &
  .0279±.0006 &
  \multicolumn{1}{c|}{.0264±.0008} &
  .0116±.0005 &
  .0163±.0004 &
  .0148±.0005 \\
  \multicolumn{1}{l|}{ReCAFR+SimGCL} &
  \textbf{.0241}±.0007 &
  \textbf{.0254}±.0005 &
  \multicolumn{1}{c|}{\textbf{.0279}±.0007} &
  \textbf{.0171}±.0001 &
  \textbf{.0261}±.0005 &
  \multicolumn{1}{c|}{\textbf{.0295}±.0004} &
  \textbf{.0274}±.0008 &
  \textbf{.0289}±.0009 &
  \multicolumn{1}{c|}{\textbf{.0329}±.0006} &
  \textbf{.0121}±.0005 &
  \textbf{.0175}±.0005 &
  \textbf{.0184}±.0001 \\ \hline
\hline
\end{tabular}%
}\label{tab:review-aware1}
\end{table*}

\begin{table*}[t]
\centering
\caption{Performance comparison (@20) between ReCAFR and review-based baselines. The best result in each setting is bolded.}
\vspace{-2mm}
\addtolength{\tabcolsep}{-1mm}
\renewcommand{\arraystretch}{1.2}
\resizebox{0.99\textwidth}{!}{%
\begin{tabular}{lccc|ccc|ccc|ccc}
\hline \hline
  \multicolumn{1}{c|}{\textbf{Method}} &
  \multicolumn{3}{c|}{\textbf{Kindle}} &
  \multicolumn{3}{c|}{\textbf{Book}} &
  \multicolumn{3}{c|}{\textbf{Beauty}} &
  \multicolumn{3}{c}{\textbf{Yelp}} \\ \hline 
    \multicolumn{13}{c}{\textbf{Using all available reviews}} \\ \hline 
  \multicolumn{1}{l|}{} &
  Recall@20 &
  Prec@20 &
  NDCG@20 &
  Recall@20 &
  Prec@20 &
  NDCG@20 &
  Recall@20 &
  Prec@20 &
  NDCG@20 &
  Recall@20 &
  Prec@20 &
  NDCG@20 \\ \hline
  \multicolumn{1}{l|}{NARRE} &
  .0650±.0007 &
  .0172±.0003 &
  \multicolumn{1}{c|}{.0447±.0006} &
  .0496±.0003 &
  .0205±.0005 &
  \multicolumn{1}{c|}{.0384±.0002} &
  .0806±.0003 &
  .0210±.0008 &
  \multicolumn{1}{c|}{.0533±.0008} &
  .0326±.0006 &
  .0116±.0008 &
  .0241±.0006 \\
  \multicolumn{1}{l|}{RGCL} &
  .0641±.0008 &
  .0185±.0008 &
  \multicolumn{1}{c|}{.0455±.0008} &
  .0532±.0002 &
  .0221±.0002 &
  \multicolumn{1}{c|}{.0422±.0004} &
  .0812±.0002 &
  .0211±.0002 &
  \multicolumn{1}{c|}{.0537±.0000} &
  .0331±.0007 &
  .0142±.0005 &
  .0268±.0008 \\
  \multicolumn{1}{l|}{RMCL} &
  .0582±.0007 &
  .0172±.0001 &
  \multicolumn{1}{c|}{.0341±.0007} &
  .0541±.0007 &
  .0225±.0007 &
  \multicolumn{1}{c|}{.0429±.0005} &
  .0824±.0003 &
  .0206±.0003 &
  \multicolumn{1}{c|}{.0541±.0004} &
  .0291±.0006 &
  .0137±.0005 &
  .0241±.0000 \\
  \multicolumn{1}{l|}{ReCAFR+SimGCL} &
  \textbf{.0699}±.0005 &
  \textbf{.0206}±.0001 &
  \multicolumn{1}{c|}{\textbf{.0476}±.0003} &
  \textbf{.0582}±.0004 &
  \textbf{.0240}±.0008 &
  \multicolumn{1}{c|}{\textbf{.0440}±.0006} &
  \textbf{.0839}±.0004 &
  \textbf{.0214}±.0006 &
  \multicolumn{1}{c|}{\textbf{.0564}±.0008} &
  \textbf{.0349}±.0002 &
  \textbf{.0153}±.0006 &
  \textbf{.0277}±.0008 \\ \hline \hline
\multicolumn{13}{c}{\textbf{Removing 30\% reviews}} \\ \hline 
  \multicolumn{1}{l|}{} &
  Recall@20 &
  Prec@20 &
  NDCG@20 &
  Recall@20 &
  Prec@20 &
  NDCG@20 &
  Recall@20 &
  Prec@20 &
  NDCG@20 &
  Recall@20 &
  Prec@20 &
  NDCG@20 \\\hline
  \multicolumn{1}{l|}{NARRE} &
  .0601±.0009 &
  .0148±.0005 &
  \multicolumn{1}{c|}{.0406±.0009} &
  .0452±.0006 &
  .0182±.0009 &
  \multicolumn{1}{c|}{.0341±.0004} &
  .0788±.0007 &
  .0186±.0001 &
  \multicolumn{1}{c|}{.0501±.0003} &
  .0297±.0005 &
  .0101±.0001 &
  .0219±.0007 \\
  \multicolumn{1}{l|}{RGCL} &
  .0594±.0002 &
  .0154±.0006 &
  \multicolumn{1}{c|}{.0425±.0009} &
  .0501±.0003 &
  .0198±.0006 &
  \multicolumn{1}{c|}{.0376±.0003} &
  .0786±.0001 &
  .0189±.0002 &
  \multicolumn{1}{c|}{.0498±.0006} &
  .0308±.0002 &
  .0116±.0009 &
  .0227±.0004 \\
  \multicolumn{1}{l|}{RMCL} &
  .0551±.0001 &
  .0142±.0008 &
  \multicolumn{1}{c|}{.0319±.0003} &
  .0505±.0007 &
  .0204±.0006 &
  \multicolumn{1}{c|}{.0388±.0008} &
  .0773±.0009 &
  .0174±.0001 &
  \multicolumn{1}{c|}{.0413±.0001} &
  .0264±.0009 &
  .0109±.0004 &
  .0216±.0002 \\
  \multicolumn{1}{l|}{ReCAFR+SimGCL} &
  \textbf{.0673}±.0003 &
  \textbf{.0189}±.0006 &
  \multicolumn{1}{c|}{\textbf{.0454}±.0003} &
  \textbf{.0558}±.0003 &
  \textbf{.0221}±.0009 &
  \multicolumn{1}{c|}{\textbf{.0419}±.0001} &
  \textbf{.0803}±.0002 &
  \textbf{.0195}±.0006 &
  \multicolumn{1}{c|}{\textbf{.0530}±.0001} &
  \textbf{.0331}±.0005 &
  \textbf{.0124}±.0009 &
  \textbf{.0252}±.0006 \\ \hline \hline
\end{tabular}%
}\label{tab:review-aware-top20}
\end{table*}

\section{Details of baseline settings} \label{app:baseline}
We provide detailed settings for the baselines.

For all baseline models, the embedding size is fixed to $64$ and the embedding parameters (user, item, and review representations) are initialized with the Xavier method. We use a learning rate of $10^{-3}$ and the default mini-batch size of 1024 to optimize the model by the Adam optimizer. 

For the collaborative filtering approaches: The $L_2$ regularization coefficient is set to $10^{-3}$ for BPR and $10^{-4}$ for LightGCN. The number of layers in LightGCN is set to $2$. In DirectAU, hyper-parameters are tuned in the ranges suggested by the original paper except for the weight of $l_{\text{uniform}}$ is tuned from $0.01$ to $10$.
In SGL, we employ the SGL-ED variant, with the self-supervised coefficient, regularization coefficient and edge dropout coefficient set to 0.5, 0.001, and 0.1, respectively. The number of layers in SGL is set to $2$. To fine-tune hyperparameters of SimGCL, we search $L_2$ regularization from $10^{-5}$ to $10^{-2}$ and we empirically let the temperature $\tau = 0.2$ and this $\tau $ also is reported in those papers as the best.

For the review-based methods: For NARRE, the number of neurons in the convolutional layer is $100$ with window size set to $3$, and the dropout ratio is set to 0.4 to prevent overfitting. The hyper-parameters of RGCL are set to $\alpha = 0.8$ and $\beta=0.4$, to control the strength of node and edge discrimination. We set the node dropout ratio to $0.4$ in RGCL. In the experiments of RMCL, the size of latent vector is searched in \{$32,64,128$\} along with the number of intentions tuned in \{$5,10,15,20,50$\}.


\section{Details of LLM-based experiments} \label{app:LLMexp}
We describe the instruction and input prompt used in our LLM enhancement. 
Fig.~\ref{fig:LLMex} showcases an example specific to the Book dataset, following the design in RLMRec \cite{RLMRec}. Given a domain-customized instruction template, and an input prompt populated by each review, along with the title and description of the corresponding item, the LLM generates a response containing ``summarization'' and ``reasoning'' sections, representing an enhanced version of the review.


\begin{table*}[t]
\centering
\caption{Performance comparison between ReCAFR and baselines that do not use review data. Each baseline X is compared with ReCAFR+X, with the best result in each pairing bolded. }
\vspace{-2mm}
\addtolength{\tabcolsep}{-1mm}
\renewcommand{\arraystretch}{1.2}
\resizebox{1\textwidth}{!}{%
\begin{tabular}{r|ccc|ccc|ccc|ccc}
\hline \hline
  \multicolumn{1}{l|}{\textbf{}} &
  \multicolumn{3}{c|}{\textbf{Kindle}} &
  \multicolumn{3}{c|}{\textbf{Book}} &
  \multicolumn{3}{c|}{\textbf{Beauty}} &
  \multicolumn{3}{c}{\textbf{Yelp}} \\ \hline 
  \textbf{Methods} &
  Recall@5 &
  Prec@5 &
  NDCG@5 &
  Recall@5 &
  Prec@5 &
  NDCG@5 &
  Recall@5 &
  Prec@5 &
  NDCG@5 &
  Recall@5 &
  Prec@5 &
  NDCG@5 \\ \hline
\rowcolor{verylightgray}
  BPR &
  .0203±.0003 &
  .0223±.0008 &
  .0242±.0005 &
  .0142±.0009 &
  .0227±.0006 &
  .0236±.0001 &
  .0264±.0007 &
  .0272±.0009 &
  .0310±.0003 &
  .0109±.0006 &
  .0157±.0007 &
  .0166±.0002 \\
 \rowcolor{verylightgray}
  ReCAFR+BPR &
  \textbf{.0226}±.0001 &
  \textbf{.0241}±.0009 &
  \textbf{.0245}±.0001 &
  \textbf{.0143}±.0002 &
  \textbf{.0228}±.0001 &
  \textbf{.0237}±.0008 &
  \textbf{.0277}±.0003 &
  \textbf{.0287}±.0002 &
  \textbf{.0313}±.0008 &
  \textbf{.0115}±.0008 &
  \textbf{.0162}±.0004 &
  \textbf{.0169}±.0004 \\
  LightGCN &
  .0232±.0001 &
  .0253±.0007 &
  .0281±.0001 &
  .0152±.0005 &
  .0244±.0001 &
  .0254±.0006 &
  .0254±.0008 &
  .0261±.0005 &
  .0298±.0000 &
  .0112±.0006 &
  .0159±.0004 &
  .0168±.0002 \\
  ReCAFR+LightGCN &
  \textbf{.0243}±.0005 &
  \textbf{.0265}±.0009 &
  \textbf{.0293}±.0001 &
  \textbf{.0187}±.0005 &
  \textbf{.0299}±.0002 &
  \textbf{.0311}±.0001 &
  \textbf{.0266}±.0006 &
  \textbf{.0276}±.0004 &
  \textbf{.0300}±.0006 &
  \textbf{.0125}±.0006 &
  \textbf{.0168}±.0002 &
  \textbf{.0175±.0005} \\
 \rowcolor{verylightgray}
  SGL &
  .0237±.0006 &
  \textbf{.0259}±.0007 &
  .0286±.0007 &
  .0169±.0001 &
  .0270±.0008 &
  .0281±.0005 &
  .0269±.0002 &
  \textbf{.0276}±.0009 &
  .0316±.0008 &
  .0101±.0004 &
  .0143±.0001 &
  .0151±.0004 \\
 \rowcolor{verylightgray}
  ReCAFR+SGL &
  \textbf{.0242}±.0009 &
  .0257±.0005 &
  \textbf{.0291}±.0005 &
  \textbf{.0171}±.0006 &
  \textbf{.0274}±.0007 &
  \textbf{.0285}±.0007 &
  \textbf{.0276}±.0007 &
  .0273±.0004 &
  \textbf{.0319}±.0009 &
  \textbf{.0106}±.0001 &
  \textbf{.0148}±.0008 &
  \textbf{.0158}±.0001 \\ 
  DirectAU &
  .0255±.0009 &
  .0271±.0003 &
  .0309±.0007 &
  .0167±.0009 &
  .0269±.0007 &
  .0281±.0009 &
  .0298±.0007 &
  .0284±.0007 &
  .0338±.0006 &
  \textbf{.0124}±.0009 &
  .0161±.0008 &
  .0169±.0005 \\
  ReCAFR+DirectAU &
  \textbf{.0262}±.0000 &
  \textbf{.0273}±.0006 &
  \textbf{.0317}±.0009 &
  \textbf{.0172}±.0006 &
  \textbf{.0271}±.0006 &
  \textbf{.0285}±.0007 &
  \textbf{.0301}±.0008 &
  \textbf{.0286}±.0008 &
  \textbf{.0341}±.0007 &
  .0121±.0006 &
  \textbf{.0179}±.0001 &
  \textbf{.0172}±.0008 \\
   \rowcolor{verylightgray}
  SimGCL &
  .0253±.0002 &
  .0277±.0004 &
  \textbf{.0306}±.0003 &
  .0180±.0007 &
  .0289±.0006 &
  .0301±.0007 &
  .0288±.0000 &
  .0295±.0002 &
  .0339±.0007 &
  .0145±.0001 &
  .0181±.0002 &
  .0188±.0002 \\
 \rowcolor{verylightgray}
  ReCAFR+SimGCL &
  \textbf{.0269}±.0002 &
  \textbf{.0285}±.0001 &
  .0302±.0003 &
  \textbf{.0184}±.0003 &
  \textbf{.0295}±.0006 &
  \textbf{.0307}±.0001 &
  \textbf{.0296}±.0007 &
  \textbf{.0304}±.0005 &
  \textbf{.0365}±.0001 &
  \textbf{.0155}±.0007 &
  \textbf{.0191}±.0009 &
  \textbf{.0202}±.0004 \\  
  \hline \hline
\textbf{Methods} &
  Recall@20 &
  Prec@20 &
  NDCG@20 &
  Recall@20 &
  Prec@20 &
  NDCG@20 &
  Recall@20 &
  Prec@20 &
  NDCG@20 &
  Recall@20 &
  Prec@20 &
  NDCG@20 \\ \hline 
  BPR &
  \textbf{.0571}±.0001 &
  .0165±.0008 &
  .0394±.0001 &
  .0439±.0009 &
  \textbf{.0227}±.0005 &
  .0340±.0001 &
  .0750±.0009 &
  .0199±.0001 &
  .0516±.0005 &
  .0299±.0007 &
  .0109±.0003 &
  .0234±.0006 \\
  ReCAFR+BPR &
  .0553±.0009 &
  \textbf{.0174}±.0008 &
  \textbf{.0399}±.0009 &
  \textbf{.0440}±.0004 &
  .0194±.0009 &
  \textbf{.0351}±.0006 &
  \textbf{.0821}±.0003 &
  \textbf{.0213}±.0007 &
  \textbf{.0542}±.0002 &
  \textbf{.0315}±.0002 &
  \textbf{.0112}±.0005 &
  \textbf{.0241}±.0007 \\
 \rowcolor{verylightgray}
  LightGCN &
  .0616±.0002 &
  .0178±.0000 &
  .0437±.0007 &
  .0472±.0006 &
  .0195±.0008 &
  .0365±.0006 &
  .0720±.0002 &
  .0191±.0003 &
  .0496±.0006 &
  .0304±.0001 &
  .0110±.0009 &
  .0237±.0005 \\
 \rowcolor{verylightgray}
  ReCAFR+LightGCN &
  \textbf{.0645}±.0009 &
  \textbf{.0186}±.0009 &
  \textbf{.0457}±.0003 &
  \textbf{.0543}±.0002 &
  \textbf{.0225}±.0005 &
  \textbf{.0430}±.0003 &
  \textbf{.0788}±.0007 &
  \textbf{.0204}±.0009 &
  \textbf{.0520}±.0004 &
  \textbf{.0316}±.0000 &
  \textbf{.0128}±.0003 &
  \textbf{.0241}±.0008 \\
  SGL &
  .0628±.0006 &
  .0182±.0001 &
  .0446±.0001 &
  .0523±.0004 &
  .0216±.0009 &
  .0404±.0001 &
  \textbf{.0763}±.0002 &
  \textbf{.0202}±.0005 &
  \textbf{.0525}±.0009 &
  .0273±.0004 &
  .0099±.0004 &
  \textbf{.0213}±.0006 \\
  ReCAFR+SGL &
  \textbf{.0629}±.0006 &
  \textbf{.0184}±.0005 &
  \textbf{.0448}±.0003 &
  \textbf{.0525}±.0009 &
  \textbf{.0219}±.0001 &
  \textbf{.0412}±.0001 &
  .0761±.0002 &
  .0199±.0007 &
  .0513±.0008 &
  \textbf{.0278±.0003} &
  \textbf{.0107±.0001} &
  .0211±.0005 \\ 
  \rowcolor{verylightgray}
  DirectAU &
  \textbf{.0681}±.0007 &
  .0196±.0006 &
  .0481±.0007 &
  \textbf{.0529}±.0005 &
  .0254±.0009 &
  .0411±.0007 &
  .0793±.0007 &
  .0216±.0002 &
  .0546±.0006 &
  \textbf{.0307}±.0003 &
  \textbf{.0116}±.0008 &
  \textbf{.0238}±.0007 \\
 \rowcolor{verylightgray}
  ReCAFR+DirectAU &
  .0676±.0004 &
  \textbf{.0199}±.0006 &
  \textbf{.0491}±.0002 &
  .0516±.0005 &
  \textbf{.0256}±.0004 &
  \textbf{.0418}±.0005 &
  \textbf{.0801}±.0005 &
  \textbf{.0221}±.0001 &
  \textbf{.0549}±.0003 &
  .0302±.0003 &
  .0115±.0003 &
  .0223±.0001 \\
  SimGCL &
  .0659±.0006 &
  .0191±.0000 &
  .0468±.0006 &
  .0549±.0006 &
  .0227±.0008 &
  .0424±.0001 &
  .0801±.0008 &
  .0211±.0008 &
  .0551±.0004 &
  .0340±.0007 &
  .0143±.0007 &
  .0266±.0007 \\
  ReCAFR+SimGCL &
  \textbf{.0670}±.0003 &
  \textbf{.0205}±.0003 &
  \textbf{.0476}±.0009 &
  \textbf{.0582}±.0002 &
  \textbf{.0240}±.0008 &
  \textbf{.0440}±.0001 &
  \textbf{.0839}±.0001 &
  \textbf{.0214}±.0001 &
  \textbf{.0564}±.0005 &
  \textbf{.0349}±.0006 &
  \textbf{.0153}±.0003 &
  \textbf{.0277}±.0009 \\
  \hline \hline
\end{tabular}%
}
\label{tab:overall1}
\end{table*}

\section{Additional Results}

We present the performance comparison with review-based baselines for top-5 and top-20 recommendations in Table ~\ref{tab:review-aware1} and Table~\ref{tab:review-aware-top20}.

We further describe our full overall performance of ReCAFR and five collaborative baselines in Table ~\ref{tab:overall1}